\documentclass[aps,onecolumn,showpacs,nofootinbib,showkeys]{revtex4-2}
\usepackage[margin=3.15cm]{geometry}

\RequirePackage[T1]{fontenc}

\usepackage{epsfig,graphicx,amsmath,amssymb,bm}
\RequirePackage{mathptmx}      
\usepackage{mathrsfs}
\usepackage[english]{babel}
\usepackage{amssymb}
\RequirePackage{color}
\usepackage{changes}
\usepackage{slashed}
\usepackage{multirow}
\usepackage{scalerel}
\usepackage{tikz-feynman}
\usepackage{textcomp}
\usepackage{subcaption}
\usepackage[english]{babel}
\usepackage{float}
\RequirePackage{hyperref}
\hypersetup{
    linktocpage,
    colorlinks,
    citecolor=brown,
    filecolor=black,
    linkcolor=brown,
    urlcolor=purple,
}
\usepackage{nccmath}

\begin{document}

\title{Description of the decay $\tau \to K \pi \pi \nu_\tau$ in the NJL type chiral quark model}


\author{M.K. Volkov$^{1}$}\email{volkov@theor.jinr.ru}
\author{A.A. Pivovarov$^{1}$}\email{tex$\_$k@mail.ru}
\author{K. Nurlan$^{1,2,3}$}\email{nurlan@theor.jinr.ru}

\affiliation{$^1$ Bogoliubov Laboratory of Theoretical Physics, JINR, 
                 141980 Dubna, Moscow region, Russia \\
                $^2$ The Institute of Nuclear Physics, Almaty, 050032, Kazakhstan \\
                $^3$ L.N. Gumilyov Eurasian National University, Astana, 010008, Kazakhstan}   


\begin{abstract}
The branching fractions of the decays $\tau^- \to K^- \pi^+ \pi^- \nu_\tau$, $\tau^- \to K^- \pi^0 \pi^0 \nu_\tau$ and $\tau^- \to \bar{K}^0 \pi^- \pi^0\nu_\tau$ are calculated in the $U(3)\times U(3)$ Nambu--Jona-Lasinio type chiral model. Four intermediate channels are considered: the contact channel and the channels with the intermediate axial vector, vector and pseudoscalar mesons. The additional resonance states $K^* \pi$ and $K \rho$ are taken into account in all channels. It is shown that the main contributions to the widths of these decays are given by the axial vector channel. In the axial vector channel, the intermediate resonances $K_1(1270)$ and $K_1(1400)$ are taken into account. The obtained results are in satisfactory agreement with the known experimental data. 


\end{abstract}

\pacs{}

\maketitle


\section{\label{Intro}Introduction}
Numerous $\tau$ lepton decays are a good laboratory for studying of the meson interactions at low energies. One of the convenient methods for studying these decays is the Nambu-Jona-Lasinio (NJL) type phenomenological model \cite{Nambu:1961tp, Eguchi:1976iz, Volkov:1986zb, Ebert:1985kz, Vogl:1991qt, Klevansky:1992qe, Hatsuda:1994pi, Ebert:1994mf, Volkov:2005kw}. Using one of the $U(3)\times U(3)$ versions of the chiral quark model proposed by one of the authors in the early 1980s \cite{Ebert:1982pk, Volkov:1984kq, Volkov:1986zb, Volkov:1993jw, Ebert:1994mf, Volkov:2005kw}, numerous $\tau$ decays with the production of one or two mesons in the final state are successfully described \cite{Volkov:2017arr, Volkov:2022jfr}. This version of the NJL model takes into account the possibility of nondiagonal transitions between pseudoscalar and axial-vector mesons. Also, when determining the internal parameters of the model, normalization is carried out by the weak decay constant $F_{\pi} = 92.4$ MeV and the strong decay constant $g_{\rho} = 6.0$~ \cite{Gounaris:1968mw, Volkov:2021fmi}.

In addition to $\tau$ lepton decays into one or two mesons, tau decays with the production of three pseudoscalar mesons are of great interest. Here, the most probable decay is $\tau \to 3\pi \nu_\tau$, which, in particular, in our version of the NJL model was described in \cite{Ivanov:1989qw}. The next $\tau$ lepton decays with the production of three pseudoscalar mesons are with a probability of $10^{-3}$: $\tau \to \eta\pi\pi\nu_\tau$, $\tau \to K \pi\pi\nu_\tau$ and $\tau \to K K \pi\nu_\tau$. The decay of $\tau \to \eta\pi\pi\nu_\tau$ was considered within the NJL model in \cite{Volkov:2013zba}.

The present paper is devoted to the calculation of the branching fractions of $\tau^- \to K^- \pi^+ \pi^- \nu_\tau$, $\tau^- \to K^- \pi^0 \pi^0 \nu_\tau$ and $\tau^- \to \bar{K}^0 \pi^- \pi^0\nu_\tau$. These decays are interesting due to the existence of many intermediate channels with different numbers of intermediate resonances. Here the axial vector channel with two resonances $K_1(1270)$ and $K_1(1400)$ plays a decisive role.
These two states are the result of mixing the $K_{1A}$ and $K_{1B}$ states. In our model, the result of such mixing was described in \cite{Volkov:1984gqw, Volkov:2019awd, Suzuki:1993yc, Kang:2018jzg}.

As for experimental studies, the first ones of decays $\tau^- \to K^- \pi^+ \pi^- \nu_\tau$ were carried out in the late 1990s \cite{CLEO:1998cuo}. The CLEO and OPAL collaborations experiments in the early 2000s measured the decay branching fractions $Br(\tau^- \to K^- \pi^+ \pi^- \nu_\tau)= (3.84 \pm 0.38)\times 10 ^{-3}$ \cite{CLEO:2003dfk} and $Br(\tau^- \to K^- \pi^+ \pi^- \nu_\tau)= (4.15 \pm 0.53)\times 10^ {-3}$ \cite{OPAL:2004icu}. Later widths measured by the BaBar collaboration \cite{BaBar:2007chl} turned out to be significantly lower than the values of a previous measurements $Br(\tau^- \to K^- \pi^+ \pi^- \nu_\tau)= (2.73 \pm 0.09)\times 10^{-3}$. At the same time, the Belle collaboration presented the value $Br(\tau^- \to K^- \pi^+ \pi^- \nu_\tau)= (3.30 \pm 0.17)\times 10^{-3 }$ \cite{Belle:2010fal}. A scatter is observed in the measured central values of the decay branching fractions. The average value in the Particle Data Group (PDG) for this decay is $Br(\tau^- \to K^- \pi^+ \pi^- \nu_\tau)= (3.45 \pm 0.07)\times 10^{ -3}$ \cite{ParticleDataGroup:2022pth}. This discrepancy between the experimental data for the charged decay mode makes it interesting to carry out theoretical studies of such decays. At the same time, the experiments carried out by the Belle collaboration in 2014 for the decay of $\tau^- \to \bar{K}^0 \pi^- \pi^0 \nu_\tau$ \cite{Belle:2014mfl} confirmed the old results obtained in \cite{ALEPH:1997trn, ALEPH:1999jxs}.
       
\section{Lagrangian of the NJL model}
In the version of the NJL model applied in the present work, the fragment of the quark-meson interaction Lagrangian containing the needed vertices takes the form~\cite{Volkov:2005kw,Volkov:2022jfr}:
\begin{eqnarray}
	\Delta L_{int} & = & \bar{q}\left\{\sum_{i=0,\pm}\left[ig_{\pi}\gamma^{5}\lambda^{\pi}_{i}\pi^{i} +
	ig_{K}\gamma^{5}\lambda^{K}_{i}K^{i} + \frac{g_{\rho}}{2}\gamma^{\mu}\lambda^{\rho}_{i}\rho^{i}_{\mu} + \frac{g_{K^{*}}}{2}\gamma^{\mu}\lambda^{K}_{i}K^{*i}_{\mu} \right.\right. \nonumber\\
	&&\left.\left. + \frac{g_{K_1}}{2}\gamma^{\mu}\lambda^{K}_{i}K^{i}_{1A\mu}\right] + ig_{K}\gamma^{5}\lambda_{0}^{\bar{K}}\bar{K}^{0} + \frac{g_{K^{*}}}{2}\gamma^{\mu}\lambda^{\bar{K}}_{0}\bar{K}^{*0}_{\mu}\right\}q,
\end{eqnarray}
where $q$ and $\bar{q}$ are the u-, d- and s-quark fields with the constituent masses $m_{u} = m_{d} = 270$ MeV,
$m_{s} = 420$~MeV, and $\lambda$ are the linear combinations of the Gell-Mann matrices.

The axial vector meson $K_{1A}$ can be represented as a sum of two physical states~\cite{Volkov:2019awd}:
\begin{eqnarray}
    K_{1A} = K_1(1270)\sin{\alpha} + K_1(1400)\cos{\alpha},
\end{eqnarray}
where $\alpha = 57^\circ$.

The coupling constants:
\begin{displaymath}
	g_{\pi} = \sqrt{\frac{Z_{\pi}}{4 I_{20}}}, \quad g_{\rho} = \sqrt{\frac{3}{2 I_{20}}}, \quad g_{K} = \sqrt{\frac{Z_{K}}{4 I_{11}}}, \quad g_{K^{*}} = g_{K_1} = \sqrt{\frac{3}{2 I_{11}}},
\end{displaymath}
where
\begin{eqnarray}
	&Z_{\pi} = \left(1 - 6\frac{m^{2}_{u}}{M^{2}_{a_{1}}}\right)^{-1}, \quad
	Z_{K} = \left(1 - \frac{3}{2}\frac{(m_{u} + m_{s})^{2}}{M^{2}_{K_{1A}}}\right)^{-1},& \nonumber\\
	&M^{2}_{K_{1A}} = \left(\frac{\sin^{2}{\alpha}}{M^{2}_{K_{1}(1270)}} - \frac{\cos^{2}{\alpha}}{M^{2}_{K_{1}(1400)}}\right)^{-1},&
\end{eqnarray}
$Z_{\pi}$ and $Z_{K}$ are the factors appearing when pseudoscalar -- axial-vector transitions are taken into account, $M_{a_{1}} = 1230$~MeV, $M_{K_{1}(1270)} = 1253$~MeV, $M_{K_{1}(1400)} = 1403$~MeV are the masses of the axial vector mesons $a_{1}$ and $K_{1}$~\cite{ParticleDataGroup:2022pth}. Taking into consideration these transitions distinguishes the present version of the NJL model from many others. The integrals included in the definitions of the coupling constants take the following form:
\begin{equation}
\label{integral}
	I_{nm} = -i\frac{N_{c}}{(2\pi)^{4}}\int\frac{\theta(\Lambda^{2} + k^2)}{(m_{u}^{2} - k^2)^{n}(m_{s}^{2} - k^2)^{m}}
	\mathrm{d}^{4}k,
\end{equation}
where $\Lambda = 1265$~MeV is the cut-off parameter~\cite{Volkov:2022jfr}.

\section{The amplitudes of the decays $\tau \to K\pi\pi\nu_{\tau}$}
\begin{figure*}[t]
 \centering
 \centering
 \begin{subfigure}{0.5\textwidth}
  \centering
   \begin{tikzpicture}
    \begin{feynman}
      \vertex (a) {\(\tau\)};
      \vertex [dot, right=2cm of a] (b){};
      \vertex [above right=2cm of b] (c) {\(\nu_{\tau}\)};
      \vertex [dot, below right=1.2cm of b] (d) {};
      \vertex [dot, right=0.8cm of d] (l) {};
      \vertex [dot, right=1.5cm of l] (g) {};
      \vertex [dot, above right=1.2cm of g] (e) {};
      \vertex [dot, below right=1.2cm of g] (h) {};      
      \vertex [dot, right=1.2cm of e] (f) {};
      \vertex [dot, above right=1.0cm of f] (n) {};
      \vertex [dot, below right=1.0cm of f] (m) {};
      \vertex [right=1.0cm of n] (s) {\( K (\pi) \)};
      \vertex [right=1.0cm of m] (r) {\( \pi \)};
      \vertex [right=1.2cm of h] (k) {\( \pi (K) \)}; 
      \diagram* {
         (a) -- [fermion] (b),
         (b) -- [fermion] (c),
         (b) -- [boson, edge label'=\(W\)] (d),
         (d) -- [fermion, inner sep=1pt, half left] (l),
         (l) -- [fermion, inner sep=1pt, half left] (d),
         (l) -- [edge label'=\({ K_{1A}, K^{*}, K } \)] (g),
         (g) -- [anti fermion] (h),  
         (h) -- [anti fermion] (e),
         (e) -- [anti fermion] (g),      
         (e) -- [edge label'=\( K^{*} (\rho) \)] (f),
         (f) -- [fermion] (n),
         (n) -- [fermion] (m),
         (m) -- [fermion] (f),
         (h) -- [] (k),
         (n) -- [] (s),
         (m) -- [] (r),
      };
     \end{feynman}
    \end{tikzpicture}
  \end{subfigure}%
 \caption{The diagrams with the intermediate mesons describing the decays $\tau \to K^* \pi, K \rho \to K \pi \pi \nu_\tau$.}
 \label{diagram2}
\end{figure*}
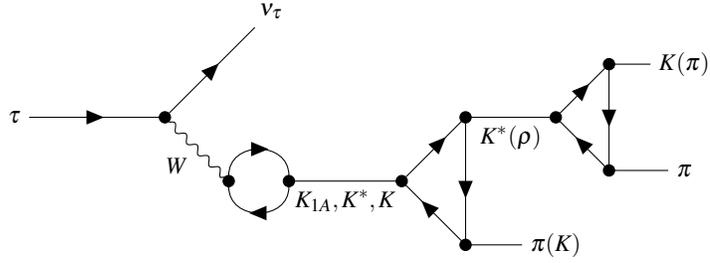

The diagrams describing the decays $\tau \to K\pi\pi\nu_{\tau}$ are shown in Figs.~\ref{diagram1} and \ref{diagram2}.

\begin{figure*}[t]
 \centering
  \begin{subfigure}{0.5\textwidth}
   \centering
    \begin{tikzpicture}
     \begin{feynman}
      \vertex (a) {\(\tau\)};
      \vertex [dot, right=2cm of a] (b){};
      \vertex [above right=2cm of b] (c) {\(\nu_{\tau}\)};
      \vertex [dot, below right=1.2cm of b] (d) {};
      \vertex [dot, above right=1.2cm of d] (e) {};
      \vertex [dot, below right=1.2cm of d] (h) {};
      \vertex [dot, right=1.2cm of e] (f) {};
      \vertex [dot, above right=1.0cm of f] (n) {};  
      \vertex [dot, below right=1.0cm of f] (m) {};   
      \vertex [right=1.2cm of n] (l) {\(\ K (\pi) \)}; 
      \vertex [right=1.2cm of m] (s) {\(\pi\)};  
      \vertex [right=1.4cm of h] (k) {\(\pi (K) \)}; 
      \diagram* {
         (a) -- [fermion] (b),
         (b) -- [fermion] (c),
         (b) -- [boson, edge label'=\(W\)] (d),
         (d) -- [fermion] (e),  
         (e) -- [fermion] (h),
         (d) -- [anti fermion] (h),
         (e) -- [edge label'=\({ K^{*} (\rho)} \)] (f),
         (f) -- [fermion] (n),
         (n) -- [fermion] (m),
         (f) -- [anti fermion] (m), 
         (h) -- [] (k),
         (n) -- [] (l),
	 (m) -- [] (s),
      };
     \end{feynman}
    \end{tikzpicture}
  \end{subfigure}%
 \caption{The contact quark diagram of the decays $\tau \to K^* \pi, K \rho \to K \pi \pi \nu_\tau$.}
 \label{diagram1}
\end{figure*}
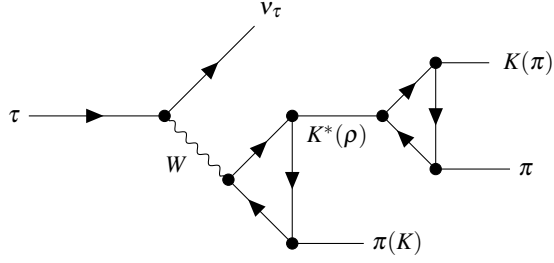%

The amplitude of all the considered processes include the axial vector, vector and pseudoscalar channels containing the axial vector, vector and pseudoscalar mesons as the first resonances, respectively:
\begin{equation}
\label{amplitude}
    \mathcal{M} = G_{F} V_{us} L_{\mu} \left\{\mathcal{M}_{A} + \mathcal{M}_{V} + \mathcal{M}_{P}\right\}^{\mu},
\end{equation}
where $L_{\mu}$ is the lepton current.

In the considered processes, the box diagrams have not been taken into account because their contributions are small and are in the framework of the model uncertainty.

\subsection{ The process $\tau \to \bar{K}^{0}\pi^{-}\pi^{0}\nu_{\tau}$}
In the process $\tau \to \bar{K}^{0}\pi^{-}\pi^{0}\nu_{\tau}$, each of the channels in the amplitude (\ref{amplitude}) is the sum of channels with the mesons $\rho^{-}$, $K^{*-}$ and $\bar{K}^{*0}$ as the second resonances:
\begin{eqnarray}
\label{contributions}
    \mathcal{M}_{A}^{\mu} & = & i\frac{3}{2\sqrt{2}} Z_{K} \frac{g_{\pi}^2}{g_{K}} \left[BW_{K_1(1270)} \sin^2{\alpha} + BW_{K_1(1400)} \cos^2{\alpha}\right]\left[g^{\mu\nu}h_{A} - q^{\mu}q^{\nu}\right] \nonumber\\
    && \times\left\{\frac{(m_{s} + m_{u})}{Z_{\pi}}BW_{\rho^-}\left(p_{\pi^-} - p_{\pi^0}\right)_{\nu} + m_{s}BW_{\bar{K}^{*0}}\left(\frac{p_{K}}{Z_{K_{1}}} - \frac{p_{\pi^0}}{Z_{a_{1}}}\right)_{\nu}+ m_{s}BW_{K^{*-}}\left(\frac{p_{\pi^-}}{Z_{a_{1}}} - \frac{p_{K}}{Z_{K_{1}}}\right)_{\nu}\right\}, \nonumber\\
    \mathcal{M}_{V}^{\mu} & = & 6\sqrt{2} g_{K}g_{\pi}^{2} I_{c} h_{V} BW_{K^{*}} \left\{\frac{4}{Z_{\pi}}BW_{\rho} + \left(\frac{1}{Z_{a_{1}}} + \frac{1}{Z_{K_{1}}}\right) BW_{\bar{K}^{*0}} + \left(\frac{1}{Z_{a_{1}}} + \frac{1}{Z_{K_{1}}}\right) BW_{K^{*-}} \right\} \nonumber\\
    && \times e^{\mu \nu \lambda \delta} p_{K\nu} p_{\pi^{-}\lambda} p_{\pi^{0}\delta}, \nonumber\\
    \mathcal{M}_{P}^{\mu} & = & i \frac{3\sqrt{2}}{4} Z_{K} \frac{g_{\pi}^2}{g_{K}} (m_{u} + m_{s}) BW_{K^-} q^{\mu} \left\{\frac{4}{Z_{K}Z_{\pi}}BW_{\rho^-}p_{K}^{\nu}\left(p_{\pi^-} - p_{\pi^0}\right)_{\nu} \right. \nonumber\\
    && \left.+ BW_{\bar{K}^{*0}}\left(\frac{p_{\pi^-}}{Z_{a_{1}}} + \frac{q}{Z_{K_{1}}}\right)^{\nu}\left(\frac{p_{K}}{Z_{K_{1}}} - \frac{p_{\pi^0}}{Z_{a_{1}}}\right)_{\nu} + BW_{K^{*-}}\left(\frac{p_{\pi^0}}{Z_{a_{1}}} + \frac{q}{Z_{K_{1}}}\right)^{\nu}\left(\frac{p_{\pi^-}}{Z_{a_{1}}} - \frac{p_{K}}{Z_{K_{1}}}\right)_{\nu}\right\},
\end{eqnarray}
where $p_{\pi^{0}}$, $p_{\pi^{-}}$ and $p_{K}$ are the momenta of the final mesons, $q$ is the monentum of the first intermediate meson and the values
\begin{eqnarray}
    Z_{a_{1}} = \left(1 - 3\frac{m_{u}(3m_{u} - m_{s})}{M_{a_{1}}^{2}}\right)^{-1}, \quad Z_{K_{1}} = \left(1 - 3\frac{m_{s}(m_{u} + m_{s})}{M_{K_{1A}}^{2}}\right)^{-1}
\end{eqnarray}
describe $\pi - a_{1}$ and $K - K_{1}$ transitions of the external mesons. The factors $h_{A}$ and $h_{V}$ take the form
\begin{eqnarray}
    h_{A} & = & M_{K_{1}}^{2} - i\sqrt{q^{2}}\Gamma_{K_{1}} - \frac{3}{2}(m_{s} + m_{u})^{2}, \nonumber\\
    h_{V} & = & M_{K^{*}}^{2} - i\sqrt{q^{2}}\Gamma_{K^{*}} - \frac{3}{2}(m_{s} - m_{u})^{2}.
\end{eqnarray}

The intermediate resonances are described by the Breit-Wigner propagators:
\begin{eqnarray}
    BW_{M} = \frac{1}{M_{M}^{2} - p^{2} - i\sqrt{p^{2}}\Gamma_{M}},
\end{eqnarray}
where $M$ designates a meson, $M_{M}$, $\Gamma_{M}$ and $p$ are its mass, width and momentum respectively .

In the vector channel, a combination of convergent integrals appears:
\begin{eqnarray}
    I_c = m_{u}\left[I_{21} + m_{u}(m_{s} - m_{u})I_{31}\right],
\end{eqnarray}
where $I_{21}$ and $I_{31}$ take the form (\ref{integral}) except the fact that they don't contain the cut-off parameter due to the convergence.

The amplitudes for the axial vector and vector channels presented in (\ref{contributions}) include appropriate contact contributions.

\subsection{The process $\tau \to K^{-}\pi^{+}\pi^{-}\nu_{\tau}$}
The process $\tau \to K^{-}\pi^{+}\pi^{-}\nu_{\tau}$, in contrast to the previous one contains only two mesons as the second resonance: $\rho^0$ and $\bar{K}^{*0}$. Appropriate contributions to the axial-vector, vector and pseudoscalar channels take the form
\begin{eqnarray}
    \mathcal{M}_{A}^{\mu} & = & i\frac{3}{2} Z_{K} \frac{g_{\pi}^2}{g_{K}} \left[BW_{K_1(1270)} \sin^2{\alpha} + BW_{K_1(1400)} \cos^2{\alpha}\right]\left[g^{\mu\nu}h_{A} - q^{\mu}q^{\nu}\right] \nonumber\\
    && \times\left\{\frac{(m_{s} + m_{u})}{Z_{\pi}}BW_{\rho^0}\left(p_{\pi^+} - p_{\pi^-}\right)_{\nu} + 2m_{s}BW_{\bar{K}^{*0}}\left(\frac{p_{\pi^+}}{Z_{a_{1}}} - \frac{p_{K}}{Z_{K_{1}}}\right)_{\nu}\right\}, \nonumber\\
    \mathcal{M}_{V}^{\mu} & = & 12 g_{K}g_{\pi}^{2} I_{c} h_{V} BW_{K^{*-}} \left\{\frac{2}{Z_{\pi}}BW_{\rho^0} + \left(\frac{1}{Z_{a_{1}}} + \frac{1}{Z_{K_{1}}}\right) BW_{\bar{K}^{*0}} \right\} \nonumber\\
    && \times e^{\mu \nu \lambda \delta} p_{K\nu} p_{\pi^{+}\lambda} p_{\pi^{-}\delta}, \nonumber\\
    \mathcal{M}_{P}^{\mu} & = & i \frac{3}{2} Z_{K} \frac{g_{\pi}^2}{g_{K}} (m_{u} + m_{s}) BW_{K^-} q^{\mu} \left\{\frac{2}{Z_{K}Z_{\pi}}BW_{\rho^0}p_{K}^{\nu}\left(p_{\pi^+} - p_{\pi^-}\right)_{\nu} \right. \nonumber\\
    && \left.+ BW_{\bar{K}^{*0}}\left(\frac{p_{\pi^-}}{Z_{a_{1}}} + \frac{q}{Z_{K_{1}}}\right)^{\nu}\left(\frac{p_{\pi^+}}{Z_{a_{1}}} - \frac{p_{K}}{Z_{K_{1}}}\right)_{\nu}\right\},
\end{eqnarray}
where $p_{\pi^{+}}$, $p_{\pi^{-}}$ and $p_{K}$ are the momenta of the final mesons.

\subsection{The process $\tau \to K^{-}\pi^{0}\pi^{0}\nu_{\tau}$}
The process $\tau \to K^{-}\pi^{0}\pi^{0}\nu_{\tau}$ differs from the previous ones by the inclusion of only one meson $K^{*-}$ as the second resonance in the intermediate state and the necessity to take into account the particle identity. The contributions of the axial vector, vector and pseudoscalar channels take the form:
\begin{eqnarray}
    \mathcal{M}_{A}^{\mu} & = & i\frac{3}{2} Z_{K} \frac{g_{\pi}^2}{g_{K}} \left[BW_{K_1(1270)} \sin^2{\alpha} + BW_{K_1(1400)} \cos^2{\alpha}\right]\left[g^{\mu\nu}h_{A} - q^{\mu}q^{\nu}\right] \nonumber\\
    && \times m_{s}BW_{K^{*-}}\left(\frac{p_{K}}{Z_{K_{1}}} - \frac{p_{\pi^0}^{(1)}}{Z_{a_{1}}}\right)_{\nu} + \left(p_{\pi^0}^{(1)} \leftrightarrow p_{\pi^0}^{(2)}\right), \nonumber\\
    \mathcal{M}_{V}^{\mu} & = & 6 g_{K}g_{\pi}^{2} I_{c} h_{V} BW_{K^{*-}} \left(\frac{1}{Z_{a_{1}}} + \frac{1}{Z_{K_{1}}}\right) BW_{K^{*-}} e^{\mu \nu \lambda \delta} p_{K\nu} p_{\pi^{0}\lambda}^{(2)} p_{\pi^{0}\delta}^{(1)} \nonumber\\
    && +  \left(p_{\pi^0}^{(1)} \leftrightarrow p_{\pi^0}^{(2)}\right), \nonumber\\
    \mathcal{M}_{P}^{\mu} & = & i \frac{3}{4} Z_{K} \frac{g_{\pi}^2}{g_{k}} (m_{u} + m_{s}) BW_{K^-} BW_{K^{*-}}q^{\mu}\left(\frac{p_{\pi^0}^{(2)}}{Z_{a_{1}}} + \frac{q}{Z_{K_{1}}}\right)^{\nu}\left(\frac{p_{\pi^0}^{(1)}}{Z_{a_{1}}} - \frac{p_{K}}{Z_{K_{1}}}\right)_{\nu} \nonumber\\
    && + \left(p_{\pi^0}^{(1)} \leftrightarrow p_{\pi^0}^{(2)}\right),
\end{eqnarray}
where $p_{\pi^0}^{(1)}$, $p_{\pi^0}^{(2)}$ and $p_{K}$ are the momenta of the final mesons.

	\begin{table}
		\caption{The model predictions of the partial widths of the decays $\tau \to K \pi\pi \nu_\tau$. The contributions of different channels are shown on different lines. The line $K_{1A}$ corresponds to the axial vector channel. The line Total contains the summed results of all channels.}
		\label{tab:1}       
		\begin{center}
		\begin{tabular}{cccccc}
			\hline\noalign{\smallskip}
			&\multicolumn{4}{c}{Br $(\times 10^{-3})$}\\
			\noalign{\smallskip}\hline\noalign{\smallskip}
			&\multicolumn{2}{c}{$\tau^- \to \bar{K}^0 \pi^- \pi^0 \nu_{\tau}$}&\multicolumn{2}{c}{$\tau^- \to K^- \pi^+ \pi^- \nu_{\tau}$} & $\tau^- \to K^- 2\pi^0 \nu_{\tau}$\\
			\noalign{\smallskip}\hline\noalign{\smallskip}
		Intermediate states	& $(K^* \pi)^-$ & $ (K\rho)^-$ & $(K^* \pi)^-$ & $ (K\rho)^-$ & $(K^* \pi)^-$ \\
			\noalign{\smallskip}\hline\noalign{\smallskip}
			$K_{1A}$	 	 & 2.66 		& 0.88		       & 2.56	      & 0.43   		& 0.68 \\
			$K^*$        	 & 0.12 		& 0.034 		   & 0.063		  & 0.017   	& 0.013  \\
			$K$   	     	 & 0.024 		& 0.028 		   & 0.021		  & 0.009   	& 0.006  \\	
			\noalign{\smallskip}\hline\noalign{\smallskip}
			Total      	 & \multicolumn{2}{c}{$ 3. 70 \pm 0.55$} & \multicolumn{2}{c}{$ 3. 27 \pm 0.49$} & $0.68 \pm 0.10$ \\
            \noalign{\smallskip}\hline\noalign{\smallskip}
            Experiment & \multicolumn{2}{c}{$ 3. 82 \pm 0.13$} \cite{ParticleDataGroup:2022pth} & \multicolumn{2}{c}{$ 3. 45 \pm 0.07$} \cite{ParticleDataGroup:2022pth} & $0.65 \pm 0.22$ \cite{ParticleDataGroup:2022pth} \\
            & \multicolumn{2}{c}{$ 3. 86 \pm 0.14$} \cite{Belle:2014mfl} & \multicolumn{2}{c}{$ 3.30 \pm 0.17$} \cite{Belle:2010fal} & \\
            & \multicolumn{2}{c}{} & \multicolumn{2}{c}{$ 2. 73 \pm 0.09$} \cite{BaBar:2007chl} \\
			\noalign{\smallskip}\hline
		\end{tabular}
		\end{center}
	\end{table}

The numerical estimations of the obtained results are presented in Table~\ref{tab:1}.

\section{Conclusion}
In the present paper, within our version of the NJL quark model, four particle $\tau$ lepton decays with the production of one K meson and two pions in the final state are described. The calculation results show that the dominant contribution in determining the decay branching fractions is given by the axial vector channel. In the axial vector channel, two physical axial vector states $K_{1}(1270)$ and $K_{1}(1400)$ are taken into account, which are the result of mixing the states $K_{1A}$ and $K_{1B}$. The contribution from the pseudoscalar channel is small and only interferes with the axial vector channel. The vector channel makes a small contribution of the order of $10^{-5}$. The amplitude of the vector channel is orthogonal and does not interfere with other channels. The axial vector and vector channels also include the corresponding contact diagrams. Here we did not take into account box type diagrams due to a negligible contribution. For example, the box diagram for the pseudoscalar channel of the process $\tau \to K^- \pi^+ \pi^- \nu_\tau$ gives the result $Br(\tau^- \to K^- \pi^+ \pi^ - \nu_\tau)_{box} = 2.6 \times 10^{-6}$. The results of all contributions obtained in our version of the NJL model are presented in Table~\ref{tab:1}. The model precision is estimated as $\pm15\%$ based on the statistical analysis of numerous previous calculations and partial axial current conservation (PCAC) \cite{Volkov:2022jfr}. 
 
Interesting results are obtained for the decay of the $\tau$ lepton into charged mesons $\tau \to K^- \pi^+ \pi^- \nu_\tau$, where there is a discrepancy between the experimentally measured decay widths. Our results are in good agreement with the latest data from the Belle collaboration \cite{Belle:2010fal}. It is important to note that our results are obtained without using any additional arbitrary parameters.

The $\tau$ lepton decays into three pseudoscalars containing one K meson were described in \cite{Decker:1992kj, Finkemeier:1995sr} from a theoretical point of view using chiral Lagrangians, and the following estimates for the branching fractions were obtained: $Br(\tau^- \to K^- \pi^+ \pi^- \nu_\tau) = 5.67 \times 10^{-3}$, $Br(\tau^- \to K^- 2\pi^0 \nu_\tau) = 1.08 \times 10^{-3}$, $Br(\tau^- \to \bar{K}^0 \pi^- \pi^0 \nu_\tau) = 5.77 \times 10^{-3}$ \cite{Decker:1992kj} and $Br(\tau^- \to K^- \pi^+ \pi^- \nu_\tau) = 7.70 \times 10^{-3}$, $Br(\tau^- \to K^- 2\pi^0 \nu_\tau) = 1.40 \times 10^{-3}$, $Br(\tau^- \to \bar{K}^0 \pi^- \pi^0 \nu_\tau) = 9.60 \times 10^{-3}$ \cite{Finkemeier:1995sr}. Unfortunately, these results noticeably diverge from experimental data.

\subsection*{Acknowledgments}
The authors are grateful to prof. A.B. Arbuzov for useful discussions.



\begin{thebibliography}{99}
\bibitem{Nambu:1961tp} 
  Y.~Nambu and G.~Jona-Lasinio,
  Dynamical Model of Elementary Particles Based on an Analogy with Superconductivity. 1.,
  Phys.\ Rev.\  {\bf 122}, 345 (1961).
  doi:10.1103/PhysRev.122.345
  
\bibitem{Eguchi:1976iz} 
  T.~Eguchi,
  A New Approach to Collective Phenomena in Superconductivity Models,
  Phys.\ Rev.\ D {\bf 14}, 2755 (1976).
  doi:10.1103/PhysRevD.14.2755

\bibitem{Volkov:1986zb} 
  M.~K.~Volkov,
  Low-energy Meson Physics in the Quark Model of Superconductivity Type. (In Russian),
  Sov.\ J.\ Part.\ Nucl.\  {\bf 17}, 186 (1986)
  [Fiz.\ Elem.\ Chast.\ Atom.\ Yadra {\bf 17}, 433 (1986)].
  
\bibitem{Ebert:1985kz} 
  D.~Ebert and H.~Reinhardt,
  Effective Chiral Hadron Lagrangian with Anomalies and Skyrme Terms from Quark Flavor Dynamics,
  Nucl.\ Phys.\ B {\bf 271}, 188 (1986).
  doi:10.1016/0550-3213(86)90359-7, 10.1016/S0550-3213(86)80009-8
  
\bibitem{Vogl:1991qt} 
  U.~Vogl and W.~Weise,
  The Nambu and Jona Lasinio model: Its implications for hadrons and nuclei,
  Prog.\ Part.\ Nucl.\ Phys.\  {\bf 27}, 195 (1991).
  doi:10.1016/0146-6410(91)90005-9
  
\bibitem{Klevansky:1992qe} 
  S.~P.~Klevansky,
  The Nambu-Jona-Lasinio model of quantum chromodynamics,
  Rev.\ Mod.\ Phys.\  {\bf 64}, 649 (1992).
  doi:10.1103/RevModPhys.64.649

\bibitem{Hatsuda:1994pi} 
  T.~Hatsuda and T.~Kunihiro,
  QCD phenomenology based on a chiral effective Lagrangian,
  Phys.\ Rept.\  {\bf 247}, 221 (1994)
  doi:10.1016/0370-1573(94)90022-1
  [hep-ph/9401310].
  
\bibitem{Ebert:1994mf} 
  D.~Ebert, H.~Reinhardt and M.~K.~Volkov,
  Effective hadron theory of QCD,
  Prog.\ Part.\ Nucl.\ Phys.\  {\bf 33}, 1 (1994).
  doi:10.1016/0146-6410(94)90043-4
  
\bibitem{Volkov:2005kw} 
  M.~K.~Volkov and A.~E.~Radzhabov,
  The Nambu-Jona-Lasinio model and its development,
  Phys.\ Usp.\  {\bf 49}, 551 (2006)
  doi:10.1070/PU2006v049n06ABEH005905
  [hep-ph/0508263].

\bibitem{Ebert:1982pk}
D.~Ebert and M.~K.~Volkov,
Z. Phys. C \textbf{16} (1983), 205
doi:10.1007/BF01571607

\bibitem{Volkov:1984kq}
M.~K.~Volkov,
Annals Phys. \textbf{157} (1984), 282-303
doi:10.1016/0003-4916(84)90055-1

\bibitem{Volkov:1993jw}
M.~K.~Volkov,
Phys. Part. Nucl. \textbf{24} (1993), 35-58

\bibitem{Volkov:2017arr}
M.~K.~Volkov and A.~B.~Arbuzov,
Phys. Usp. \textbf{60} (2017) no.7, 643-666
doi:10.3367/UFNe.2016.11.037964

\bibitem{Volkov:2022jfr}
M.~K.~Volkov, A.~A.~Pivovarov and K.~Nurlan,
Symmetry \textbf{14} (2022) no.2, 308
doi:10.3390/sym14020308
[arXiv:2201.03951 [hep-ph]].

\bibitem{Gounaris:1968mw}
G.~J.~Gounaris and J.~J.~Sakurai,
Phys. Rev. Lett. \textbf{21} (1968), 244-247
doi:10.1103/PhysRevLett.21.244

\bibitem{Volkov:2021fmi}
M.~K.~Volkov, A.~A.~Osipov, A.~A.~Pivovarov and K.~Nurlan,
Phys. Rev. D \textbf{104}, no.3, 034021 (2021)
doi:10.1103/PhysRevD.104.034021
[arXiv:2105.02160 [hep-ph]]

\bibitem{Ivanov:1989qw}
Y.~P.~Ivanov, A.~A.~Osipov and M.~K.~Volkov,
Z. Phys. C \textbf{49} (1991), 563-568
doi:10.1007/BF01483571

\bibitem{Volkov:2013zba}
M.~K.~Volkov, A.~B.~Arbuzov and D.~G.~Kostunin,
Phys. Rev. C \textbf{89} (2014) no.1, 015202
doi:10.1103/PhysRevC.89.015202
[arXiv:1310.8484 [hep-ph]].

\bibitem{Volkov:1984gqw}
M.~K.~Volkov and A.~A.~Osipov,
Sov. J. Nucl. Phys. \textbf{41} (1985), 500-503
JINR-E2-84-298.

\bibitem{Volkov:2019awd}
M.~K.~Volkov and A.~A.~Pivovarov,
Phys. Part. Nucl. Lett. \textbf{16}, no.6, 565-568 (2019)

\bibitem{Suzuki:1993yc}
M.~Suzuki,
Phys. Rev. D \textbf{47} (1993), 1252-1255
doi:10.1103/PhysRevD.47.1252

\bibitem{Kang:2018jzg}
X.~W.~Kang, T.~Luo, Y.~Zhang, L.~Y.~Dai and C.~Wang,
Eur. Phys. J. C \textbf{78} (2018) no.11, 909
doi:10.1140/epjc/s10052-018-6385-9
[arXiv:1808.02432 [hep-ph]].

\bibitem{CLEO:1998cuo}
S.~J.~Richichi \textit{et al.} [CLEO],
Phys. Rev. D \textbf{60} (1999), 112002
doi:10.1103/PhysRevD.60.112002
[arXiv:hep-ex/9810026 [hep-ex]].

\bibitem{CLEO:2003dfk}
R.~A.~Briere \textit{et al.} [CLEO],
Phys. Rev. Lett. \textbf{90} (2003), 181802
doi:10.1103/PhysRevLett.90.181802
[arXiv:hep-ex/0302028 [hep-ex]].

\bibitem{OPAL:2004icu}
G.~Abbiendi \textit{et al.} [OPAL],
Eur. Phys. J. C \textbf{35} (2004), 437-455
doi:10.1140/epjc/s2004-01877-2
[arXiv:hep-ex/0406007 [hep-ex]].

\bibitem{BaBar:2007chl}
B.~Aubert \textit{et al.} [BaBar],
Phys. Rev. Lett. \textbf{100} (2008), 011801
doi:10.1103/PhysRevLett.100.011801
[arXiv:0707.2981 [hep-ex]].

\bibitem{Belle:2010fal}
M.~J.~Lee \textit{et al.} [Belle],
Phys. Rev. D \textbf{81} (2010), 113007
doi:10.1103/PhysRevD.81.113007
[arXiv:1001.0083 [hep-ex]].

\bibitem{ParticleDataGroup:2022pth}
R.~L.~Workman \textit{et al.} [Particle Data Group],
PTEP \textbf{2022} (2022), 083C01
doi:10.1093/ptep/ptac097

\bibitem{Belle:2014mfl}
S.~Ryu \textit{et al.} [Belle],
Phys. Rev. D \textbf{89} (2014) no.7, 072009
doi:10.1103/PhysRevD.89.072009
[arXiv:1402.5213 [hep-ex]].

\bibitem{ALEPH:1997trn}
R.~Barate \textit{et al.} [ALEPH],
Eur. Phys. J. C \textbf{4} (1998), 29-45
doi:10.1007/s100529800879

\bibitem{ALEPH:1999jxs}
R.~Barate \textit{et al.} [ALEPH],
Eur. Phys. J. C \textbf{10} (1999), 1-18
doi:10.1007/s100529900146
[arXiv:hep-ex/9903014 [hep-ex]].

\bibitem{Decker:1992kj}
R.~Decker, E.~Mirkes, R.~Sauer and Z.~Was,
Z. Phys. C \textbf{58} (1993), 445-452
doi:10.1007/BF01557702

\bibitem{Finkemeier:1995sr}
M.~Finkemeier and E.~Mirkes,
Z. Phys. C \textbf{69} (1996), 243-252
doi:10.1007/s002880050024
[arXiv:hep-ph/9503474 [hep-ph]].

\end{thebibliography}
\end{document}